\documentclass{iauc}
\usepackage{graphicx}

\title[The Origin of Ultra-Compact Dwarf Galaxies in the Virgo Cluster] 
{The Origin of UCDs in the Virgo Cluster}

\author[E.A. Evstigneeva, M.D. Gregg, M.J. Drinkwater]   
{E.A. Evstigneeva$^{1}$, M.D. Gregg$^{2}$%
  \thanks{Also at: Institute for Geophysics and Planetary Physics, Lawrence Livermore
National Laboratory, L-413, Livermore, CA 94550, USA},
 \and M.J. Drinkwater$^{1}$}
 
\affiliation{$^1$Department of Physics, University of Queensland, QLD 4072, Australia \break
e-mail: katya@physics.uq.edu.au \\[\affilskip]
$^2$Department of Physics, University of California, Davis, CA 95616, USA}

\pubyear{2005}
\volume{198} 
\pagerange{1--3}
\date{?? and in revised form ??}
\jname{Near-Field Cosmology with Dwarf Elliptical Galaxies }
\editors{H. Jerjen and B. Binggeli, eds.}
\begin{document}

\maketitle

\begin{abstract}
We present the internal velocity dispersions and spectral line
indices for six Virgo ultra-compact dwarf (UCD) galaxies obtained from spectroscopy 
with the Keck II 10-m telescope. 
We use these results: 1) to compare the Virgo UCDs with the Fornax UCD population;
2) to compare UCDs with globular
clusters (GCs) to determine if UCDs and the most luminous GCs
are the same or distinct systems; 3) to further test the nucleated dwarf
elliptical (dE,N) stripping hypothesis for UCD formation.
\keywords{galaxies: dwarf, galaxies: kinematics and dynamics, galaxies: abundances, 
galaxies: clusters: individual (Virgo Cluster)}

\end{abstract}

\firstsection 
\section{Introduction}

UCDs are a new type of galaxy recently discovered in the center of 
the Fornax, Virgo, and other clusters of galaxies (\cite[Hilker et al. 1999]{hilker}; 
\cite[Drinkwater et al. 2003]{dr03},\cite[2004]{dr04}; 
\cite[Jones et al. 2005]{jones}). Possible origins of UCDs include the following hypotheses: 
1) they are luminous intra-cluster globular 
clusters; 2) they are extremely luminous star clusters formed from the amalgamation of stellar 
super-clusters that were created in galaxy interactions; 3) they are the remnant nuclei of stripped 
dwarf galaxies which have lost their outer parts in the course of tidal interaction with the galaxy 
cluster potential; 4) they are highly compact galaxies formed in the early Universe.

Here we present new results of the spectroscopic analysis 
from the Keck II telescope for six Virgo UCDs (Table 1) to 
explore and test hypotheses of their formation.

\section{Internal Dynamics}

{\it Figure~1} compares UCDs to other types of stellar systems, on the luminosity - velocity dispersion 
plane. The internal velocity dispersions for the UCDs (and GCs) were measured from CaT region 
($8400-8750\mbox{\AA}$) 
using the direct-fitting method (\cite[van der Marel 1994]{marel}). We include the most massive and 
luminous known GCs:
G1, $\omega$ Cen, and NGC5128 GCs as well as GCs in the Milky Way and M31, dE,Ns  
and their nuclei, and giant ellipticals. We find:
1) The Virgo UCDs have the same dynamical properties as the Fornax ones 
(\cite[Drinkwater et al. 2003]{dr03}).
2) There is no gap between luminous GCs and UCDs in this plot. We obtained velocity dispersions for 
the two brightest M87 GCs from \cite[Hanes et al. (2001)]{hanes} list: they lie in the same part 
of $M_V - \sigma$ plane as UCDs. 
3) The UCDs follow approximately the same relation between luminosity and velocity dispersion as GCs, 
except for the two brightest ones. More data on the velocity dispersions for bright GCs or fainter UCDs 
are required to confirm this. 
4) There is an overlap in luminosities and velocity dispersions of the nuclei of dE,Ns and the properties 
of bright GCs and UCDs . It supports the dE,N tidal stripping hypothesis for UCD and GC formation. 

\begin{table}
\begin{center}
\caption{Targets. $M_V$ -- the V band absolute magnitude. 
Strom547 and Strom417 are the two brightest globular clusters in M87 
(the central galaxy in Virgo) according to \cite[Hanes et al. (2001)]{hanes} list. 
Strom547 has luminosity comparable to that of UCDs and was therefore included into UCD list.}
\begin{tabular}{ccccc}
\hline
Name & Object type & R.A.(J2000) & Dec.(J2000) & $M_V$ \\
\hline
UCD1/Strom547 & UCD/GC & 12h30m57.40s & +12$^\circ$25$'$44.8$''$ & -12.3 \\
UCD2 & UCD & 12h30m07.61s & +12$^\circ$36$'$31.1$''$ & -12.2 \\
UCD3 & UCD & 12h31m04.51s & +11$^\circ$56$'$36.8$''$ & -12.1 \\
UCD4 & UCD & 12h31m28.41s & +12$^\circ$25$'$03.3$''$ & -12.0 \\
UCD5 & UCD & 12h31m52.93s & +12$^\circ$15$'$59.5$''$ & -13.9 \\
UCD6 & UCD & 12h31m11.90s & +12$^\circ$41$'$01.2$''$ & -12.1 \\
Strom417 & GC & 12h31m01.29s & +12$^\circ$19$'$25.6$''$ & -11.6 \\
\hline
\end{tabular}
\end{center}
\end{table}

\begin{figure}
\centering
 \resizebox{9.7cm}{!}{\includegraphics{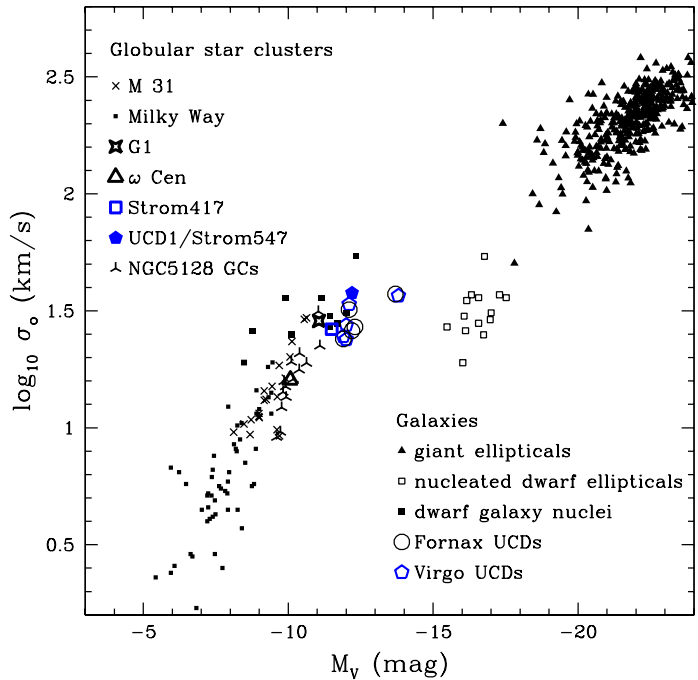} }
  \caption{Comparison of the internal dynamics of UCDs (our data) with globular clusters and 
nucleated dwarf ellipticals (literature data).}
\end{figure}

\section{Ages and Chemical Compositions}

We estimate ages, metallicities, and abundances of UCDs using Lick/IDS indices 
(\cite[Worthey et al. 1994]{worthey})  
and single-aged stellar population models of \cite[Thomas et al. (2003)]{thomas}.

In {\it Figure~2a} we plot abundance ratios of the Virgo UCDs. To estimate  [$\alpha$/Fe], 
we plot Mgb (an indicator of $\alpha$-elements) versus $<$Fe$>$ (an average of the 
indices Fe5270 and Fe5335). [$\alpha$/Fe] traces the timescale of star formation activity in 
galaxies. Most $\alpha$-elements are produced rapidly by 
Type II supernovae, while Fe is produced by Type Ia SNe on longer timescales. 
We find: 
1) Four UCDs and Strom417 (GC) have super-solar abundance ratio, 
  [$\alpha$/Fe] $\sim$ +0.3 - +0.5, and two UCDs appear to have solar abundances, 
  [$\alpha$/Fe] $\sim$ 0.0. The super-solar abundances are typical for old stellar  
  populations (in globular clusters and elliptical galaxies).
2) Virgo dE,Ns (\cite[Geha et al. 2003]{geha}) are also 
  shown in the same plot. The majority of dE,Ns are consistent with solar   
  abundance ratios, while the majority of UCDs have super-solar abundances.  
  This can be the evidence that UCDs and dE,Ns have different star formation  
  histories. 

{\it Figure~2b} shows ages and metallicities of the Virgo UCDs. We plot the age-sensitive 
H$\beta$ index versus the metallicity sensitive [MgFe]' index (\cite[Thomas et al. 2003]{thomas}).
We find:
1) The UCDs are old and metal-poor, except UCD1 which appears to be   
  intermediate-aged and metal-rich.  
2) The ages and metallicities of UCDs (except UCD1) are similar to Strom417 
  (GC) and to those found for M87 globular clusters by \cite[Cohen et al. (1998)]{cohen}.  
3) The UCDs have older integrated stellar populations than present-day dE,Ns.
  It is consistent with the stripping formation scenario, when the ISM is removed 
  from the UCD progenitor and star formation activity ceases.

\begin{figure}
\centering
\resizebox{6.7cm}{!}{\includegraphics{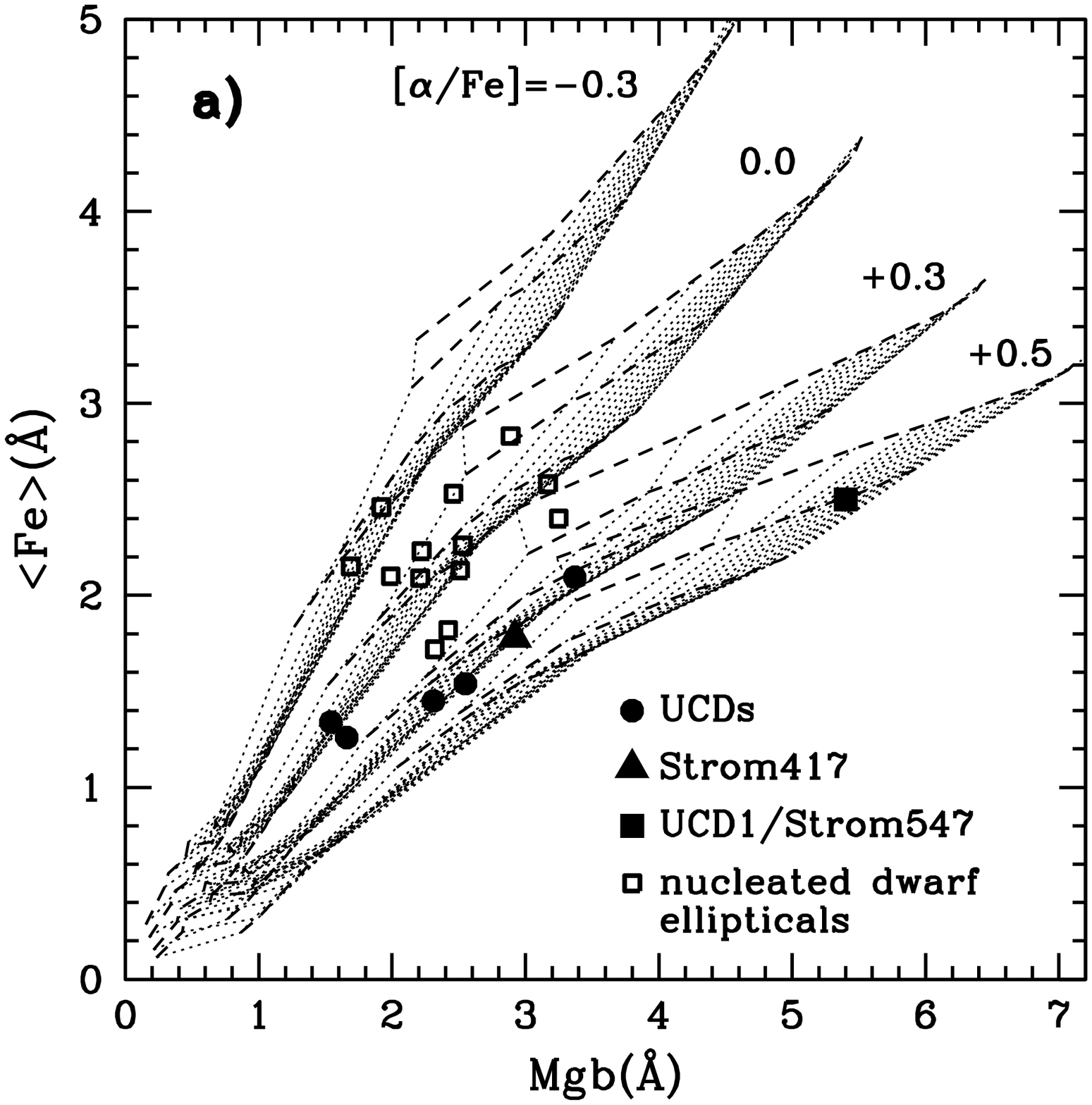} }
\resizebox{6.7cm}{!}{\includegraphics{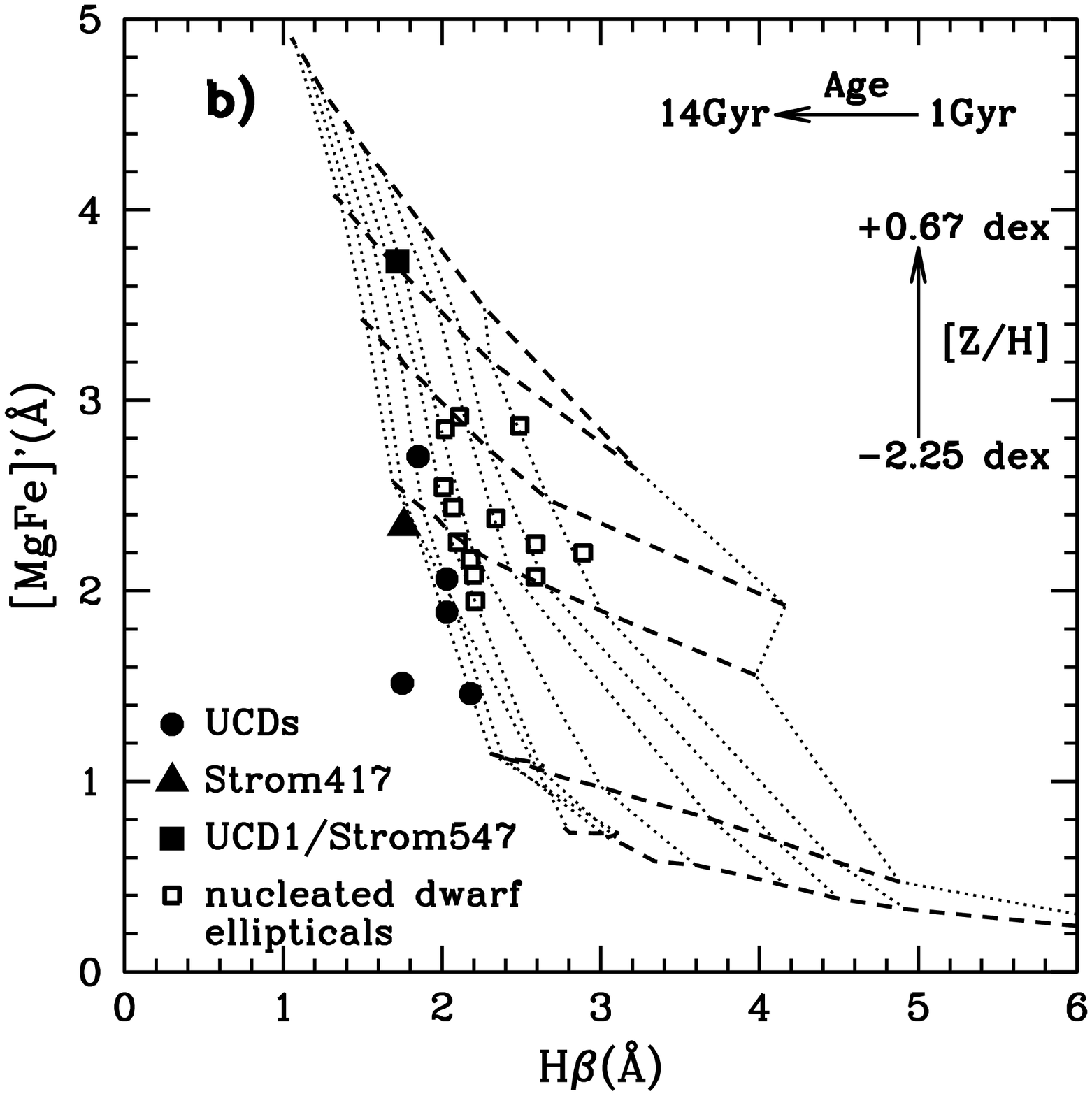} }
\caption[]{Comparison of our data with model grids from Thomas et al. (2003).\\
{\bf a)} Mgb - $<$Fe$>$.   
Thomas et al. models with variable [$\alpha$/Fe] are shown for ages 1,2,3,...,15 Gyr 
(dotted lines, from left to right) and metallicities -2.25,-1.35,-0.33,0.0,+0.35,+0.67 dex 
(dashed lines, from bottom to top). {\bf b)} H$\beta$ - [MgbFe]'.   
Thomas et al. models are shown for ages 1,2,3,4,6,8,10,12,14 Gyr 
(dotted lines, from right to left) and metallicities 
-2.25,-1.35,-0.33,0.0,+0.35,+0.67 dex (dashed lines, from bottom to top).}
\end{figure}


\end{document}